\documentclass[useAMS,usenatbib]{mn2e}
\usepackage{graphicx, rotating}
\usepackage{aas_macros}  
\usepackage{graphicx}
\usepackage{amssymb}
\usepackage{rotating}
\usepackage{multirow}
\usepackage[breaklinks,colorlinks=true,citecolor=black,linkcolor=black,urlcolor=blue,bookmarks=true]{hyperref}

\title[Emission sparks in dSph galaxies]{Emission sparks around M 81 and in some 
dSph galaxies
\thanks{Based on observations made with the 6m BTA telescope of
the Special Astrophysical Observatory, Russian Academy of Sciences.}
\thanks{Based on observations made with the NASA/ESA Hubble Space Telescope,
obtained from the data archive at the Space Telescope Science Institute.
STScI is operated by the Association of Universities for Research
in Astronomy, Inc. under NASA contract NAS 5-26555.}
}

\author[Karachentsev et al.]{
Igor Karachentsev$^1$\thanks{E-mail: ikar@sao.ru},
Elena Kaisina$^1$,
Serafim Kaisin$^1$,
Lidia Makarova$^1$ \\
$^{1}$Special Astrophysical Observatory, Nizhniy Arkhyz,
Karachai-Cherkessia 369167, Russia
}

\begin{document}

\date{Accepted XXX. Received XXX; in original form XXX}

\pagerange{\pageref{firstpage}--\pageref{lastpage}} \pubyear{XXX}

\maketitle

\label{firstpage}

\begin{abstract}
We use $H\alpha$ images of three clumps of young stars situated
between M\,81 and NGC\,3077 to estimate their star formation rate.
Radial velocities of the clumps measured by us, as well as the
velocity of HII-region in the dSph galaxy KDG\,61 are compatible
with their location at the outskirts of a large rotating gaseous
disc around M\,81. In contrast to KDG\,61, radial velocity of the
emission knot in the dSph galaxy DDO\,44, $+213\pm25$ km s$^{-1}$,
tells us that this $H\alpha$ spark belongs to the dSph galaxy
itself.
F475W and F814W images of DDO\,44 extracted from the HST archive
reveal 8 bluish ($B-I < 0.8$) stars apparently associated with the
$H\alpha$ knot.
\end{abstract}


\section{Introduction}
Systematic $H\alpha$ line observations  of the nearby ($D<10$ Mpc)
galaxies have been recently conducted at the 6-m telescope of the
Special Astrophysical Observatory, Russian Academy of Sciences 
(SAO RAS) in order to determine the rate of star formation in them
\citep{k05,kk06,kk08,kai07,kk07,kk10}. Unlike
other similar programs \citep{hun93,bk01,jam04,hun04,ken08},
we did not restrict our program to any selected
morphological types of galaxies. With almost the same enthusiasm
we observe both the gas-rich spiral, irregular and blue compact
galaxies, as well as the ``dead'' elliptical, lenticular and dwarf
spheroidal galaxies, where the current rates of star formation are
assumed to be close to zero. Such non-selective approach to the
compilation of target list has led to detection of a circumnuclear
$H\alpha$ emission from a number of isolated E, S0 galaxies 
\citep{mois10}, indicating the ongoing quasi-stationary
process of accretion of the intergalactic gas onto the central
parts of galaxies.

Another unexpected result of our survey was the discovery in some
dSph galaxies of small emission clumps, which we designated as
``sparks''. As it is known, the neighboring group around the giant
spiral M\,81 is rather rich in dwarf spheroidal systems. Some of
them: BK5N, BK6N, FM1, IKN, KKH\,70 do not show any signs of
emission in the $H\alpha$ line, and the optical bodies of
others (DDO\,44, DDO\,78, F8D1, KDG\,63) reveal fine structural details
after the subtraction of continuum {\bf \citep{kk07} }. Such details may be the result
of an incomplete subtraction of the continuum of very red stars,
or an artifact from cosmic rays. It seems unlikely that the old,
``bald'', devoid of neutral hydrogen spheroidal dwarfs contain
small sites of star formation. To verify the nature
of the hypothesized emission clumps in dSph galaxies, we carried out
spectral observations, the results of which are given in this article.

\section{Emission sparks around M\,81}

\begin{figure*}
\includegraphics[width=12cm]{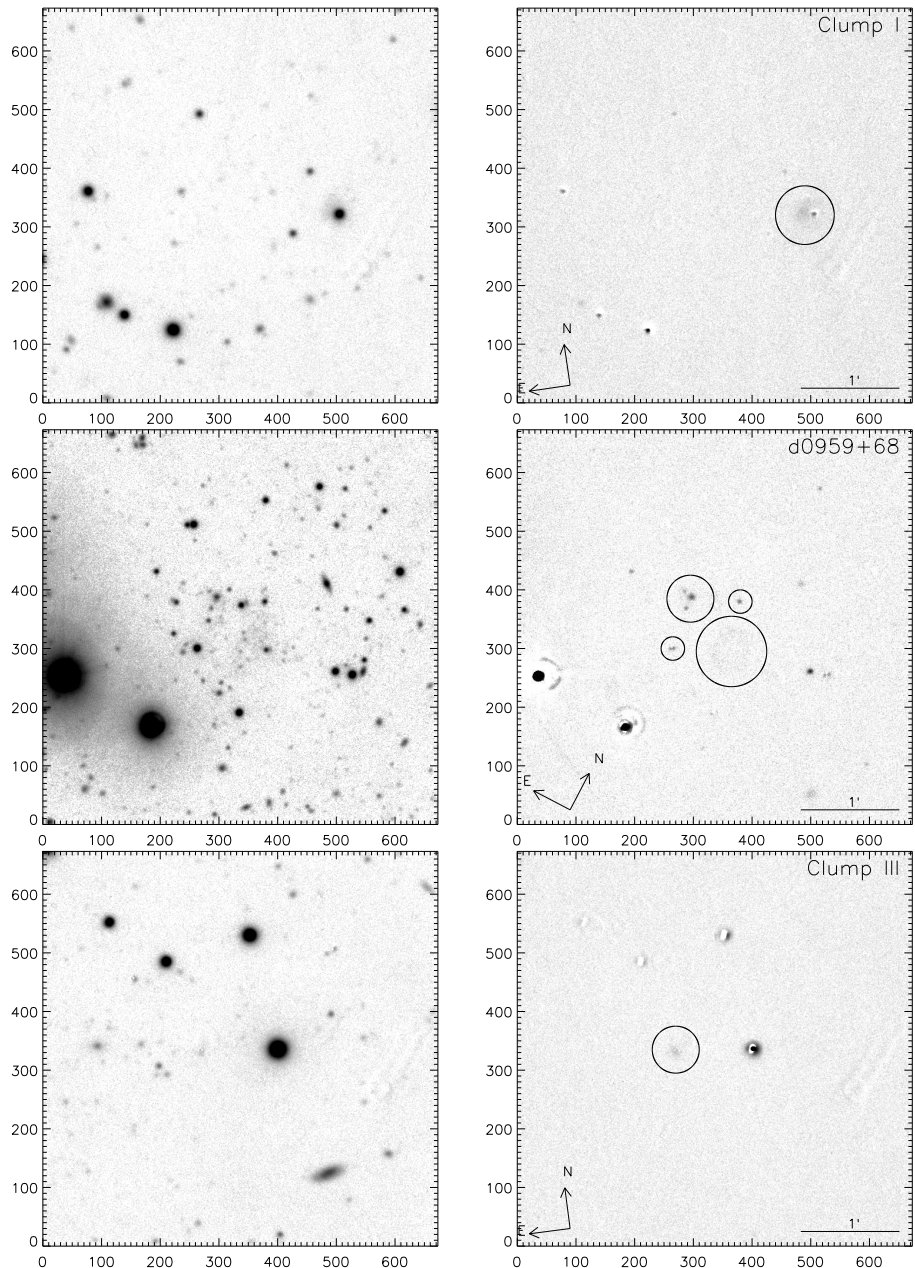}
\caption{$H\alpha$ (left) and continuum-subtracted (right) images of three
emission clumps between M\,81 and NGC\,3077. North and east are indicated by
arrows.}
\end{figure*}

As shown by \citet{app81,yun97,boyce01}, the region of the M\,81 group  
is filled with filament
structures of neutral hydrogen, which connect M\,81 with the
neighboring bright galaxies M\,82, NGC\,3077 and NGC\,2976. It is
assumed that this complex HI pattern was formed as a result of tidal
interaction of the brightest group members. In the most dense
parts of the HI filaments, the process of star formation is
already underway. It led to the formation of tidal dwarfs:
Garland, Holmberg\,IX, Arp\,loop (A0958+66) and BK3N \citep{mak02}, 
where the old ($T>2$ Gyr) stellar population is absent. \citet{bri08}
and \citet{chy11} have found in the  M\,81 group a
significant number of small HI-clouds with masses of
$\sim10^5-10^6 M_{\odot}$, free-floating between bright galaxies.
Some of them coincide in position with the dSph dwarf galaxies,
for example, KKH\,57.

\citet{mouh10} obtained deep images with  MegaCam at the
CFHT in the ``g'' and ``i'' filters of an area sized $\sim 1$
square degree between M\,81 and NGC\,3077 at subarcsecond seeing.
On these images the authors found three knots: clump I, clump II, and
clump III, resolved into blue stars. All of them are located
approximately along the HI arm, connecting M\,81 and NGC\,3077.
These bluish clumps are similar to other blue star complexes,
detected earlier in the closer periphery of M\,81
\citep{dur04,demel08,dav08}.
Note that similar groups of young (blue) stars were recently
detected with the ultraviolet GALEX satellite on the periphery of
other nearby galaxies: NGC\,404 \citep{thi10}, NGC\,628,
NGC\,2841, NGC\,3621 and NGC\,5055.

Figure~1 shows the images of the Clump I, II and III, obtained with
the 6-m BTA telescope of the SAO RAS. The  images in the $H\alpha$
line and in the continuum were made in November 2010 with the
SCORPIO focal reducer \citep{afan05}. A $2048\times2048$
pixel CCD chip  provided the field of view of 6' with the
resolution of 0.18" per pixel. To obtain the images in $H\alpha$
we used an interference filter  75\AA \-wide at the effective
wavelength of 6555\AA, while to subtract the continuum pairs of
images with the filters SED607 ($\lambda_{ef}=6063$\AA,
\,$\Delta\lambda=167$\AA) and SED707 ($\lambda_{ef}=7063$\AA,
$\Delta\lambda=207$\AA) were taken.
The exposure time amounted to $2\times600$s in $H\alpha$ and
$2\times300$s in the continuum. To calibrate the $H\alpha$ fluxes,
we also observed the spectrophotometric standards by \citet{oke99}.
The left-hand side images in Fig.~1 correspond to the total image
$H\alpha$ +  continuum, and the right-hand images show the
difference  $H\alpha$ --continuum. To determine the $H\alpha$ flux
we used a standard sequence of procedures we previously described \citep{kk10}.

\begin{table*}
 \caption{H-alpha flux and SFR for the emission sparks}
\begin{tabular}{lcccrcc} \hline
 Object   & RA  (2000.0) & Dec   & $B_T$&    $M_B$&   logF(H$\alpha$) & log[SFR]\\
\hline
Clump I   & 09 57 21.2 &+68 42 55  &19.8 &  $-$ 7.6 & $-$15.04  &   $-$4.82 \\
d0959+68  & 09 59 33.1 &+68 39 25  &18.0 &  $-$10.2 & $-$13.99  &   $-$3.77\\
Clump III & 10 00 40.4 &+68 39 37  &19.8 &  $-$ 7.5 & $-$14.34  &   $-$4.12\\
KDG 61    & 09 57 02.7 &+68 35 30  &15.2 &  $-$12.9 & $-$13.36  &   $-$3.08\\
DDO 44    & 07 34 11.3 &+66 53 10  &15.6 &  $-$12.1 & $-$15.22  &   $-$5.07\\ \hline
\end{tabular}
\end{table*}

All the three objects exhibit a weak $H\alpha$ emission, most
pronounced in the case of the object  Clump II = d0959+68 \citep{chi09}, 
where several compact HII-regions, as well as a
low-contrast diffuse emission zone are visible. The measured
integral $H\alpha$ fluxes of these objects are presented in
Table~1. We as well added there the $H\alpha$ fluxes of the
emission clumps in two other dwarf members of the group: DDO\,44 and KDG\,61,
we previously measured \citep{kk07}. The table
also lists apparent and absolute magnitudes of the objects and their
integral star formation rate $SFR(M_{\odot}\cdot
yr^{-1})=1.27\cdot10^9\cdot F_c(H\alpha)\cdot D^2$ according to
\citet{gal84}, where  $F_c(H\alpha$) is the flux
corrected for Galactic absorption, and $D$ is the distance in Mpc.

We used the detected emission clumps  to determine their radial
velocities. Spectral observations were performed in the primary
focus of the 6-m telescope with the SCORPIO focal reducer
operating in the long slit mode. The spectra of 6 objects were
obtained in the red region with  FWHM =5\AA \, and the resolution
of 0.86\AA \, per pixel. The heliocentric velocities $V_h$ of the
observed objects, measured from the $H\alpha$ line are presented
in Table 2, with indications of velocity measurement errors. The
following columns of Table 2 list:   radial velocity of the object
relative to the centroid of the Local Group, linear projected
distance from M\,81 (or from NGC\,2403 in the case DDO\,44), as well as
the position angle of the dwarf object relative to the major axis
of the main galaxy (M\,81 or NGC\,2403). The last two rows of the
table contain our new estimates of radial velocities for the dwarf
galaxy d1019+69 from the list by \citet{chi09}, as well
as UGC\,8245, earlier observed by \citet{fal99}. The galaxy
UGC\,8245 is a possible peripheral member of the M\,81 group.

\begin{figure}
\includegraphics[width=8cm]{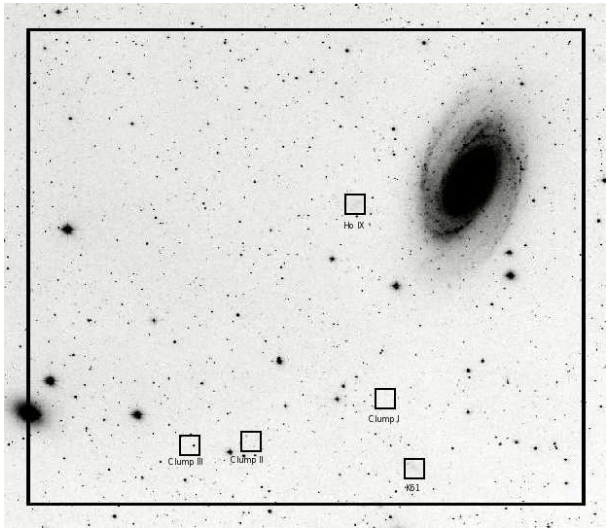}
\caption{One square degree view of the south-east outskirts of M\,81
from \citet{mouh10}. Three emission clumps and dSph galaxy
KDG\,61 are indicated by squares.}
\end{figure}

Figure~2 reproduces the  location of the observed objects relative
to the M\,81. The rectangular frame indicates the area of the image
sized  $58'\times56'$,  obtained by \citet{mouh10}. The
galaxy NGC\,3077 with the adjacent tidal dwarf structure Garland is
located on the left edge of this field. Near the lower edge of
Fig.~2 a dwarf spheroidal galaxy KDG\,61 is present. On its NE side
a bright HII region is visible, for which \citet{jon97}
have measured the radial velocity $V_h=-135\pm30$ km s$^{-1}$.
Later, similar values of radial velocity for this emission spot:
$V_h=-116\pm21$ km s$^{-1}$ and $-123\pm6$ km s$^{-1}$ were
obtained by \citet{sha01} and \citet{mak10}.
Since \citet{mak10} measured the radial velocity of
$+222\pm3$ km s$^{-1}$ for the central globular cluster in KDG\,61,
they assumed that the bright HII region is not associated with the
galaxy KDG\,61 itself, but that it is rather projected onto it from
the far periphery of the M\,81. In other words, the dSph galaxy
behaves as a screen, on which the projected structures are easily
distinguishable.

\begin{table}
\caption{New radial velocities in the M\,81 group}
\begin{tabular}{lrrrrl} \hline

Object     &   $V_h$   &   $V_{LG}$&   $R_p$  &  $\Delta PA$ &   Note\\
          &   km/s & km/s&      kpc &  deg & \\
\hline
Clump I    &  $-165\pm18$  & $- 25$ & $ 23$  & $ 7 $  &\\
d0959+68   &  $-186\pm44$  & $- 46$ & $ 35$  &  $-12$ & Clump II\\
Clump III  &  $-121\pm20$  & $  19$ & $ 39$  &  $-19$ &\\
{\bf KDG 61}  &  $-123\pm06$  & $  17$ & $ 30$  &  $ 16$ & $V_h$ from$^b$ \\
DDO 44     &  $+213\pm25$  & $ 356$ & $ 74$  &  $ 42$   &\\
d1019+69   &  $+557\pm38$  & $ 697$ &   --  & --   & backgr. dIr\\
UGC 8245   &  $-58\pm55$  &  $145$ &   --  & --   & $V_h=70\pm59^a$\\ \hline
\multicolumn{6}{l}{$^a$ \citet{fal99}}\\
\multicolumn{6}{l}{$^b$ \citet{mak10}}\\
 \hline
\end{tabular}
\end{table}

The field of HI radial velocities for the galaxy M\,81, built by \citet{rot74}
up to the angular distances of $\sim30$' from
the center shows that at the southern periphery of the M81 the
position angle sector of $\pm20^{\circ}$ relative to the
major axis of M\,81 is dominated by the radial velocities
$V_h\simeq[-140\pm30]$ km s$^{-1}$. As seen from Table~2, the
velocities of Clumps I, II, III, and emission knot in KDG\,61 lie
exactly within the specified range. Consequently, all these
emission objects can be considered as the elements of the far
periphery of the M\,81 gas disk, which has a nonuniform filamentary
structure with small sites of star formation.

\section{The emission clump in DDO\,44}
A dwarf spheroidal galaxy DDO\,44 is located at 79 arcmin in the direction
of NNW from the Sc galaxy NGC\,2403. The distance to it, based on
the tip of the red giant branch was measured by \citet{k99,alo06,dal09}. 
The mean distance value amounts to  $3.20\pm0.10$ Mpc,
which almost coincides with the distance estimate of $3.13\pm0.10$
Mpc to NGC2403 from the Cepheids \citep{fre01}. The
emission knot we discovered in DDO44 with the coordinates 073419.1
+665323.5 (J2000.0) is located in the NE side of the galaxy
within its optical boundaries.

\begin{figure}
\includegraphics[width=8cm]{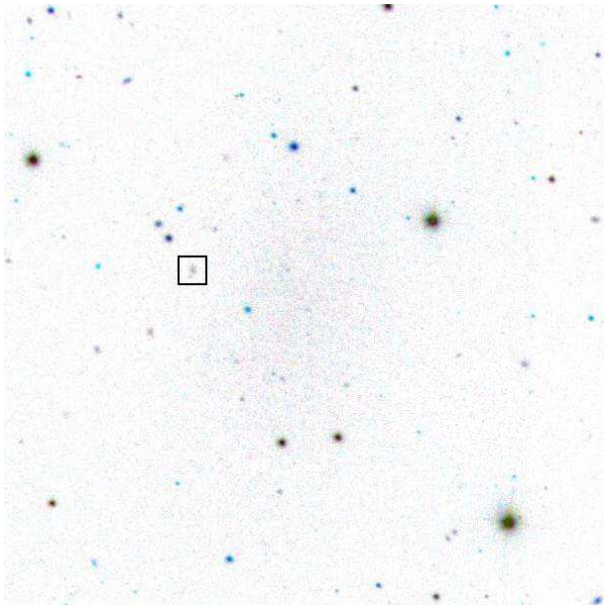}
\caption{SDSS image of DDO\,44. The emission spark is indicated by square.}
\end{figure}

On the negative image of DDO\,44, extracted from the SDSS, 
the $H\alpha$ spark is marked by a square
(Fig.~3). The heliocentric radial velocity of the emission spot in
DDO\,44 ($+213\pm25)$ km s$^{-1}$  differs from the mean velocity
of NGC\,2403 $(+133\pm3)$ km s$^{-1}$ by 80 km s$^{-1}$, which is
typical for dwarf satellites of spiral galaxies.

The field of radial velocities in NGC\,2403 was constructed by
\citet{beg87} based on observations in the neutral hydrogen
21 cm line. The resulting map of the velocity field extends to the
distance of  $\sim18$ arcmin from the center of NGC\,2403, i.e. about a
quarter of its distance to DDO\,44. From these data, the radial
velocities in the gas disk of NGC\,2403 in the direction towards DDO\,44
show a systematic decrease from the value of +133 to +60 km $^{-1}$.
Therefore, the radial velocity of the clump in DDO\,44 that we measured
in no way fits into the kinematics of the peripheral gas disk of
NGC\,2403, and can obviously be attributed to the dSph system
itself.

From the HST data archive we have picked out ACS
images of DDO\,44, taken with the F814W and F475W filters for the
GO10915 program (PI: J.~Dalcanton), and performed the DOLPHOT \citep{dol02}
photometry of stellar populations in this galaxy. A fragment of the
blue (F475W) image, sized $25$"$\times$25", with the focus on
the emission object is shown in Figure~4. Emission area with the
diameter of $\sim4$" has a rather complex spiral-like structure, due to
which this HII region can easily be confused with a distant Sc
galaxy. {\bf The results of our photometry show that some of bluish
objects with color indices of $B-I<0.8$ classified as stars based on
the DOLPHOT quality parameters are definitely associated
with the emission object.}

\begin{figure}
\includegraphics[width=8cm]{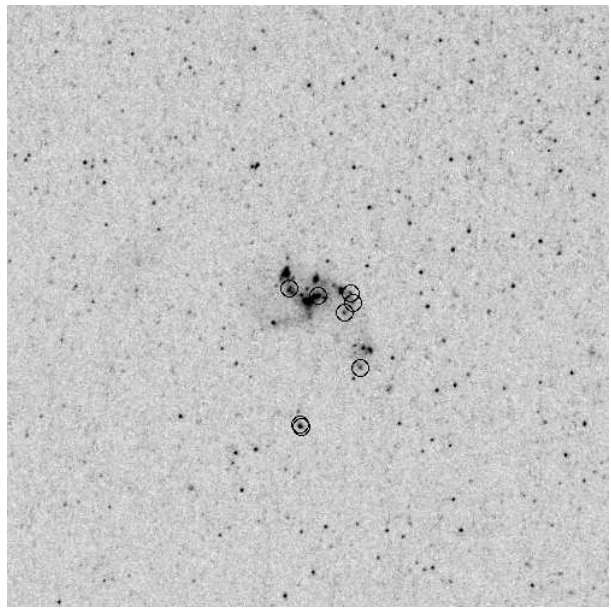}
\caption{F475W-band image of the HII-region in DDO\,44. The image size is
25 arcsec (388 pc). Eight bluish $(B-I < 0.8)$ stars are marked by circles.}
\end{figure}

Eight of the bluish stars, closest to the center of the $H\alpha$
knot are marked by circles in Figure~4. The median of the
apparent magnitudes for these 8 stars equals  $B=26.7^m$,  while
their median color index is $B-I=0.60$.  At the distance module of
$27.5^m$ the median absolute magnitude of these stars corresponds
to $M_B\simeq-1.2$. {\bf It is likely that they are late-type B-stars
with $M_{\star} > 3 M_{\sun}$, and their total far UV-luminosity, 
$ \sim 1\,10^{36} erg ^{-1}$ is
sufficient to ensure the ionization of the HII region.}

{\bf Note that the emission spark in DDO\,44 has been also detected with
the GALEX as a knot
of $m_{FUV} = 22.52 \pm0.20$ and $m_{NUV} = 21.40 \pm0.14$. Using the
relation  
$\log \textit{SFR} [M_{\sun}\,\textrm{yr}^{-1}] = 
2\,\log D [\textrm{Mpc}] - 0.4\,m_{FUV} + 2.78$
from \citet{ken98}, we obtain for the spark $\log \textit{SFR} = - 5.21$
in excellent agreement with the quantity $- 5.07$ derived
by us from the $H\alpha$ flux.}

\section{Discussion}

As follows from the above data, the emission ``sparks'' found in
the body of some nearby dwarf spheroidal galaxies have a dual
nature. They can be either compact HII-regions, projected onto the
dSph galaxy from far periphery of neighboring spiral
galaxies (the case of KDG\,61), or small sites of
star formation in dSph galaxies themselves (the case of DDO\,44).
The rate of star formation in these $H\alpha$ knots are
characterized by the values SFR$\sim(10^{-5}-10^{-3})M_{\odot}$/yr
with the integral luminosity of  $L_B\sim(1-2)\cdot10^5L_{\odot}$,
and the linear size of $\sim 50$ pc.

Another compact emission  object was discovered by us in an
isolated dSph galaxy KKR\,25 \citep{kk08}. Spectral
properties of this object are currently under study by Makarov et al.
(in preparation).

An obvious question arises: do such small emission regions exist
in closer dSph galaxies of the Local Group? It is yet difficult to
answer this question, since there were no systematic $H\alpha$
surveys of spheroidal satellites of our Galaxy conducted because
of their large angular extent.

However, in the $H\alpha$ survey of Andromeda's satellites \citet{kk06} 
noted the possible presence of faint
emission objects on the periphery of NGC\,147, And\,III and And\,X,
stressing the need to verify their nature by spectral
observations. It should be remembered here that the group
M\,81/NGC\,2403 differs from the Local Group by the presence of large
masses of intergalactic neutral hydrogen.

According to the HI observations by \citet{huc00}, the
dwarf spheroidal galaxy DDO\,44 does not reveal any HI flux at the
level of 6 mJy. Assuming the range of internal motions of
$W\leq30$ km s$^{-1}$, we obtain for DDO\,44 the upper limit of the
flux, corresponding to the hydrogen mass of
$\sim4\cdot10^5M_{\odot}$. It seems surprising how such a gas-poor
dwarf system could form a new site of star formation.

It is possible that the existence of small single $H\alpha$ clumps
in dSph galaxies is not due to internal causes, but rather to the
accretion of intergalactic gas. The accretion process can have a
universal character, but for all that be only notable  in the
``dead'' systems with old population.

Further $H\alpha$ observations of dSph galaxies in other groups,
as well as the use of ultraviolet survey data obtained by the
GALEX satellite can be of great help in studying the phenomenon of
small-scale sites of star formation in these galaxies.

\section*{Acknowledgements}
This work was supported by the RFBR grant nos. 10--02--00123 and
10--02--92650. The authors acknowledge the SDSS and GALEX database used
for this study.

\bibliographystyle{mn2e}
\bibliography{emiss}

\begin{thebibliography}{40}
\expandafter\ifx\csname natexlab\endcsname\relax\def\natexlab#1{#1}\fi

\bibitem[{{Afanasiev} \& {Moiseev}(2005)}]{afan05}
{Afanasiev} V.~L., {Moiseev} A.~V., 2005, Astronomy Letters, 31, 194

\bibitem[{{Alonso-Garc{\'{\i}}a} {et~al.}(2006){Alonso-Garc{\'{\i}}a}, {Mateo},
  \& {Aparicio}}]{alo06}
{Alonso-Garc{\'{\i}}a} J., {Mateo} M., {Aparicio} A., 2006, \pasp, 118, 580

\bibitem[{{Appleton} {et~al.}(1981){Appleton}, {Davies}, \&
  {Stephenson}}]{app81}
{Appleton} P.~N., {Davies} R.~D., {Stephenson} R.~J., 1981, \mnras, 195, 327

\bibitem[{{Begeman}(1987)}]{beg87}
{Begeman} K.~G., 1987, PhD thesis, , Kapteyn Institute, (1987)

\bibitem[{{Bell} \& {Kennicutt}(2001)}]{bk01}
{Bell} E.~F., {Kennicutt} Jr. R.~C., 2001, \apj, 548, 681

\bibitem[{{Boyce} {et~al.}(2001){Boyce}, {Minchin}, {Kilborn}, {Disney},
  {Lang}, {Jordan}, {Grossi}, {Lyne}, {Cohen}, {Morison}, \&
  {Phillipps}}]{boyce01}
{Boyce} P.~J., {Minchin} R.~F., {Kilborn} V.~A., {Disney} M.~J., {Lang} R.~H.,
  {Jordan} C.~A., {Grossi} M., {Lyne} A.~G., {Cohen} R.~J., {Morison} I.~M.,
  {Phillipps} S., 2001, \apjl, 560, L127

\bibitem[{{Brinks} {et~al.}(2008){Brinks}, {Walter}, \& {Skillman}}]{bri08}
{Brinks} E., {Walter} F., {Skillman} E.~D., 2008, in IAU Symposium, Vol. 244,
  IAU Symposium, {J.~Davies \& M.~Disney}, ed., pp. 120--126

\bibitem[{{Chiboucas} {et~al.}(2009){Chiboucas}, {Karachentsev}, \&
  {Tully}}]{chi09}
{Chiboucas} K., {Karachentsev} I.~D., {Tully} R.~B., 2009, \aj, 137, 3009

\bibitem[{{Chynoweth} {et~al.}(2011){Chynoweth}, {Langston}, \&
  {Holley-Bockelmann}}]{chy11}
{Chynoweth} K.~M., {Langston} G.~I., {Holley-Bockelmann} K., 2011, \aj, 141, 9

\bibitem[{{Dalcanton} {et~al.}(2009){Dalcanton}, {Williams}, {Seth}, {Dolphin},
  {Holtzman}, {Rosema}, {Skillman}, {Cole}, {Girardi}, {Gogarten},
  {Karachentsev}, {Olsen}, {Weisz}, {Christensen}, {Freeman}, {Gilbert},
  {Gallart}, {Harris}, {Hodge}, {de Jong}, {Karachentseva}, {Mateo}, {Stetson},
  {Tavarez}, {Zaritsky}, {Governato}, \& {Quinn}}]{dal09}
{Dalcanton} J.~J., {Williams} B.~F., {Seth} A.~C., {Dolphin} A., {Holtzman} J.,
  {Rosema} K., {Skillman} E.~D., {Cole} A., {Girardi} L., {Gogarten} S.~M.,
  {Karachentsev} I.~D., {Olsen} K., {Weisz} D., {Christensen} C., {Freeman} K.,
  {Gilbert} K., {Gallart} C., {Harris} J., {Hodge} P., {de Jong} R.~S.,
  {Karachentseva} V., {Mateo} M., {Stetson} P.~B., {Tavarez} M., {Zaritsky} D.,
  {Governato} F., {Quinn} T., 2009, \apjs, 183, 67

\bibitem[{{Davidge}(2008)}]{dav08}
{Davidge} T.~J., 2008, \pasp, 120, 1145

\bibitem[{{de Mello} {et~al.}(2008){de Mello}, {Smith}, {Sabbi}, {Gallagher},
  {Mountain}, \& {Harbeck}}]{demel08}
{de Mello} D.~F., {Smith} L.~J., {Sabbi} E., {Gallagher} J.~S., {Mountain} M.,
  {Harbeck} D.~R., 2008, \aj, 135, 548

\bibitem[{{Dolphin}(2002)}]{dol02}
{Dolphin} A.~E., 2002, \mnras, 332, 91

\bibitem[{{Durrell} {et~al.}(2004){Durrell}, {Decesar}, {Ciardullo},
  {Hurley-Keller}, \& {Feldmeier}}]{dur04}
{Durrell} P.~R., {Decesar} M.~E., {Ciardullo} R., {Hurley-Keller} D.,
  {Feldmeier} J.~J., 2004, in IAU Symposium, Vol. 217, Recycling Intergalactic
  and Interstellar Matter, {P.-A.~Duc, J.~Braine, \& E.~Brinks}, ed., pp. 90--+

\bibitem[{{Falco} {et~al.}(1999){Falco}, {Kurtz}, {Geller}, {Huchra}, {Peters},
  {Berlind}, {Mink}, {Tokarz}, \& {Elwell}}]{fal99}
{Falco} E.~E., {Kurtz} M.~J., {Geller} M.~J., {Huchra} J.~P., {Peters} J.,
  {Berlind} P., {Mink} D.~J., {Tokarz} S.~P., {Elwell} B., 1999, \pasp, 111,
  438

\bibitem[{{Freedman} {et~al.}(2001){Freedman}, {Madore}, {Gibson}, {Ferrarese},
  {Kelson}, {Sakai}, {Mould}, {Kennicutt}, {Ford}, {Graham}, {Huchra},
  {Hughes}, {Illingworth}, {Macri}, \& {Stetson}}]{fre01}
{Freedman} W.~L., {Madore} B.~F., {Gibson} B.~K., {Ferrarese} L., {Kelson}
  D.~D., {Sakai} S., {Mould} J.~R., {Kennicutt} Jr. R.~C., {Ford} H.~C.,
  {Graham} J.~A., {Huchra} J.~P., {Hughes} S.~M.~G., {Illingworth} G.~D.,
  {Macri} L.~M., {Stetson} P.~B., 2001, \apj, 553, 47

\bibitem[{{Gallagher} {et~al.}(1984){Gallagher}, {Hunter}, \&
  {Tutukov}}]{gal84}
{Gallagher} III J.~S., {Hunter} D.~A., {Tutukov} A.~V., 1984, \apj, 284, 544

\bibitem[{{Huchtmeier} {et~al.}(2000){Huchtmeier}, {Karachentsev},
  {Karachentseva}, \& {Ehle}}]{huc00}
{Huchtmeier} W.~K., {Karachentsev} I.~D., {Karachentseva} V.~E., {Ehle} M.,
  2000, \aaps, 141, 469

\bibitem[{{Hunter} \& {Elmegreen}(2004)}]{hun04}
{Hunter} D.~A., {Elmegreen} B.~G., 2004, \aj, 128, 2170

\bibitem[{{Hunter} {et~al.}(1993){Hunter}, {Hawley}, \& {Gallagher}}]{hun93}
{Hunter} D.~A., {Hawley} W.~N., {Gallagher} III J.~S., 1993, \aj, 106, 1797

\bibitem[{{James} {et~al.}(2004){James}, {Shane}, {Beckman}, {Cardwell},
  {Collins}, {Etherton}, {de Jong}, {Fathi}, {Knapen}, {Peletier}, {Percival},
  {Pollacco}, {Seigar}, {Stedman}, \& {Steele}}]{jam04}
{James} P.~A., {Shane} N.~S., {Beckman} J.~E., {Cardwell} A., {Collins} C.~A.,
  {Etherton} J., {de Jong} R.~S., {Fathi} K., {Knapen} J.~H., {Peletier} R.~F.,
  {Percival} S.~M., {Pollacco} D.~L., {Seigar} M.~S., {Stedman} S., {Steele}
  I.~A., 2004, \aap, 414, 23

\bibitem[{{Johnson} {et~al.}(1997){Johnson}, {Lawrence}, {Terlevich}, \&
  {Carter}}]{jon97}
{Johnson} R.~A., {Lawrence} A., {Terlevich} R., {Carter} D., 1997, \mnras, 287,
  333

\bibitem[{{Kaisin} \& {Karachentsev}(2006)}]{kk06}
{Kaisin} S.~S., {Karachentsev} I.~D., 2006, Astrophysics, 49, 287

\bibitem[{{Kaisin} \& {Karachentsev}(2008)}]{kk08}
---, 2008, \aap, 479, 603

\bibitem[{{Kaisin} {et~al.}(2007){Kaisin}, {Kasparova}, {Knyazev}, \&
  {Karachentsev}}]{kai07}
{Kaisin} S.~S., {Kasparova} A.~V., {Knyazev} A.~Y., {Karachentsev} I.~D., 2007,
  Astronomy Letters, 33, 283

\bibitem[{{Karachentsev} \& {Kaisin}(2007)}]{kk07}
{Karachentsev} I.~D., {Kaisin} S.~S., 2007, \aj, 133, 1883

\bibitem[{{Karachentsev} \& {Kaisin}(2010)}]{kk10}
---, 2010, \aj, 140, 1241

\bibitem[{{Karachentsev} {et~al.}(2005){Karachentsev}, {Kajsin}, {Tsvetanov},
  \& {Ford}}]{k05}
{Karachentsev} I.~D., {Kajsin} S.~S., {Tsvetanov} Z., {Ford} H., 2005, \aap,
  434, 935

\bibitem[{{Karachentsev} {et~al.}(1999){Karachentsev}, {Sharina}, {Grebel},
  {Dolphin}, {Geisler}, {Guhathakurta}, {Hodge}, {Karachentseva}, {Sarajedini},
  \& {Seitzer}}]{k99}
{Karachentsev} I.~D., {Sharina} M.~E., {Grebel} E.~K., {Dolphin} A.~E.,
  {Geisler} D., {Guhathakurta} P., {Hodge} P.~W., {Karachentseva} V.~E.,
  {Sarajedini} A., {Seitzer} P., 1999, \aap, 352, 399

\bibitem[{{Kennicutt}(1998)}]{ken98}
{Kennicutt} Jr. R.~C., 1998, \araa, 36, 189

\bibitem[{{Kennicutt} {et~al.}(2008){Kennicutt}, {Lee}, {Funes}, {Sakai}, \&
  {Akiyama}}]{ken08}
{Kennicutt} Jr. R.~C., {Lee} J.~C., {Funes} Jos{\'e}~G. S.~J., {Sakai} S.,
  {Akiyama} S., 2008, \apjs, 178, 247

\bibitem[{{Makarova} {et~al.}(2010){Makarova}, {Koleva}, {Makarov}, \&
  {Prugniel}}]{mak10}
{Makarova} L., {Koleva} M., {Makarov} D., {Prugniel} P., 2010, \mnras, 406,
  1152

\bibitem[{{Makarova} {et~al.}(2002){Makarova}, {Grebel}, {Karachentsev},
  {Dolphin}, {Karachentseva}, {Sharina}, {Geisler}, {Guhathakurta}, {Hodge},
  {Sarajedini}, \& {Seitzer}}]{mak02}
{Makarova} L.~N., {Grebel} E.~K., {Karachentsev} I.~D., {Dolphin} A.~E.,
  {Karachentseva} V.~E., {Sharina} M.~E., {Geisler} D., {Guhathakurta} P.,
  {Hodge} P.~W., {Sarajedini} A., {Seitzer} P., 2002, \aap, 396, 473

\bibitem[{{Moiseev} {et~al.}(2010){Moiseev}, {Karachentsev}, \&
  {Kaisin}}]{mois10}
{Moiseev} A., {Karachentsev} I., {Kaisin} S., 2010, \mnras, 403, 1849

\bibitem[{{Mouhcine} \& {Ibata}(2010)}]{mouh10}
{Mouhcine} M., {Ibata} R., 2010, ArXiv e-prints

\bibitem[{{Oke}(1990)}]{oke99}
{Oke} J.~B., 1990, \aj, 99, 1621

\bibitem[{{Rots} \& {Shane}(1974)}]{rot74}
{Rots} A.~H., {Shane} W.~W., 1974, \aap, 31, 245

\bibitem[{{Sharina} {et~al.}(2001){Sharina}, {Karachentsev}, \&
  {Burenkov}}]{sha01}
{Sharina} M.~E., {Karachentsev} I.~D., {Burenkov} A.~N., 2001, \aap, 380, 435

\bibitem[{{Thilker} {et~al.}(2010){Thilker}, {Bianchi}, {Schiminovich}, {Gil de
  Paz}, {Seibert}, {Madore}, {Wyder}, {Rich}, {Yi}, {Barlow}, {Conrow},
  {Forster}, {Friedman}, {Martin}, {Morrissey}, {Neff}, \& {Small}}]{thi10}
{Thilker} D.~A., {Bianchi} L., {Schiminovich} D., {Gil de Paz} A., {Seibert}
  M., {Madore} B.~F., {Wyder} T., {Rich} R.~M., {Yi} S., {Barlow} T., {Conrow}
  T., {Forster} K., {Friedman} P., {Martin} C., {Morrissey} P., {Neff} S.,
  {Small} T., 2010, \apjl, 714, L171

\bibitem[{{Yun}(1999)}]{yun97}
{Yun} M.~S., 1999, in IAU Symposium, Vol. 186, Galaxy Interactions at Low and
  High Redshift, {J.~E.~Barnes \& D.~B.~Sanders}, ed., pp. 81--+

\end{thebibliography}

\label{lastpage}
\end{document}